\documentclass[aip,jap,reprint,floatfix,amsmath,amssymb]{revtex4-1}

\usepackage{lmodern}
\usepackage{graphicx}
\usepackage{bm}
\usepackage{color}
\usepackage[percent]{overpic}
\usepackage{pdfsync}
\usepackage[breaklinks=true,colorlinks,citecolor=blue,linkcolor=blue,urlcolor=blue]{hyperref}
\usepackage[normalem]{ulem}

\begin{document}

\title{Photocurrent-based detection of Terahertz radiation in graphene}

\author{Andrea Tomadin}
\email{andrea.tomadin@sns.it}
\affiliation{NEST, Istituto Nanoscienze-CNR and Scuola Normale Superiore, I-56126 Pisa,~Italy}

\author{Alessandro Tredicucci}
\affiliation{NEST, Istituto Nanoscienze-CNR and Scuola Normale Superiore, I-56126 Pisa,~Italy}

\author{Vittorio Pellegrini}
\affiliation{Istituto Italiano di Tecnologia (IIT), Via Morego 30, 16163 Genova, Italy}
\affiliation{NEST, Istituto Nanoscienze-CNR and Scuola Normale Superiore, I-56126 Pisa,~Italy}

\author{Miriam S.~Vitiello}
\affiliation{NEST, Istituto Nanoscienze-CNR and Scuola Normale Superiore, I-56126 Pisa,~Italy}

\author{Marco Polini}
\affiliation{NEST, Istituto Nanoscienze-CNR and Scuola Normale Superiore, I-56126 Pisa,~Italy}

\begin{abstract}
Graphene is a promising candidate for the development of detectors of Terahertz (THz) radiation. A well-known detection scheme due to Dyakonov and Shur exploits the confinement of plasma waves in a field-effect transistor (FET), whereby a dc {\it photovoltage} is generated in response to a THz field. This scheme has already been experimentally studied in a graphene FET [L. Vicarelli et al.,~\href{http://dx.doi.org/10.1038/nmat3417}{Nature Mat. {\bf 11}, 865 (2012)}].
In the quest for devices with a better signal-to-noise ratio, we theoretically investigate a plasma-wave photodetector in which a dc {\it photocurrent} is generated in a graphene FET. The rectified current features a peculiar change of sign when the frequency of the incoming radiation matches an even multiple of the fundamental frequency of plasma waves in the FET channel. The noise equivalent power per unit bandwidth of our device is shown to be much smaller than that of a Dyakonov-Shur detector in a wide spectral range.
\end{abstract}

\maketitle

In a series of pioneering papers~\cite{dyakonov_prl_1993, dyakonov_prb_1995, dyakonov_ieee_1996a, dyakonov_ieee_1996b}, Dyakonov and Shur proposed a mechanism enabling detection of Terahertz (THz) radiation, which is based on the fact that a field-effect transistor (FET) hosting a two-dimensional (2D) electron gas acts as a cavity for {\it plasma waves}.
In the Dyakonov-Shur (DS) scheme plasma waves are launched by modulating the potential difference between gate and source.
When a plasma wave launched at the source can reach the drain in a time shorter than the momentum relaxation time, the detection of radiation exploits constructive interference in the cavity.
In this case one achieves frequency-resolved detection of the incoming radiation (``resonant regime").
Broadband detection occurs when plasma waves are overdamped or when the length of the FET channel is larger than the length over which a plasma wave can travel.

In the DS detection scheme a dc {\it photovoltage} is generated between source and drain in response to the incoming oscillating field.
In the resonant regime, the dc photoresponse is characterized by peaks at the odd multiples of the lowest plasma-wave frequency.
For typical device lengths and carrier densities, the fundamental plasma-wave frequency $\nu_{\rm P}$ is in the THz range, so that photodetectors based on the DS mechanism are naturally useful in the context of THz detection.
We emphasize that a substantial amount of experimental work has been carried out on DS photodetection in ordinary (III,V) semiconductors~\cite{knap_jimtw_2009,knap_nanotech_2013}. For the sake of completeness, we point out that resonant excitation and detection of 2D and 3D plasma-wave oscillations has been demonstrated also outside of the DS scheme. For example, in Ref.~\onlinecite{otsuji_apl_2004} interband photoexcitation is used to launch a plasma wave, instead of modulating the potential difference between gate and source, while in Ref.~\onlinecite{kim_apl_2008} the change in resistance between source and drain is measured, instead of a dc photovoltage.

\begin{figure}[b]
\includegraphics[width=1.0\linewidth]{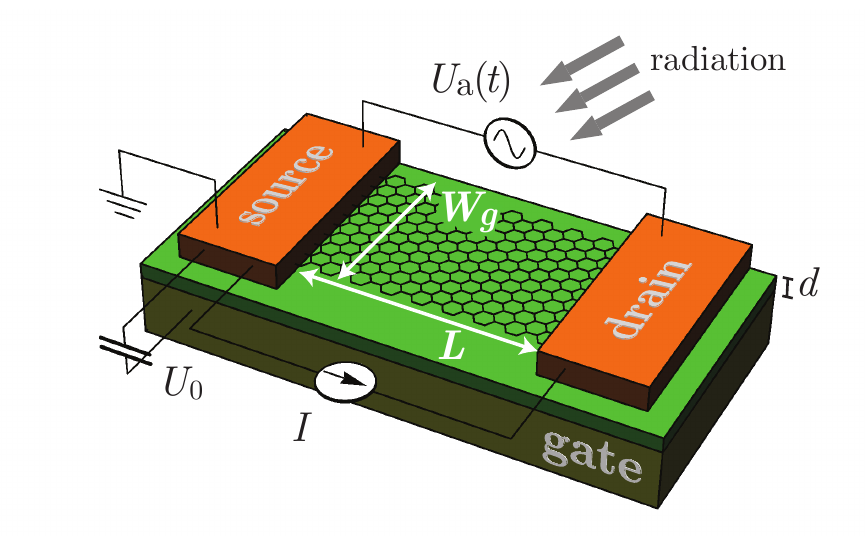}
\caption{\label{fig:setup}
Schematic representation of the setup studied in this work. The drain and source contacts of a graphene FET are connected to the feeds of an antenna (not shown), which collects impinging THz radiation.
An oscillating potential difference is generated between drain and source, which is kept at constant potential $U_{0}$ with respect to a back gate. The latter is confined in the region below the sample to avoid detrimental effects on the antenna operation. 
A dc photocurrent $I$, which is proportional to the power of the incoming radiation, is generated between source and drain. }
\end{figure}

Recently, it has been understood that graphene can pave the way for the realization of robust and cheap THz detectors operating at room temperature and based on the DS scheme~\cite{vicarelli_naturemat_2012,tredicucci_ieee_2014}.
Graphene, a 2D crystal of Carbon atoms packed in a honeycomb lattice~\cite{geim_naturemater_2007,castroneto_rmp_2009}, has indeed high carrier mobility, even at room temperature, a gapless spectrum, and a frequency-independent absorption, making it an ideal platform for a variety of applications in photonics, optoelectronics, and plasmonics~\cite{bonaccorso_naturephoton_2010,grigorenko_naturephoton_2012}.
Vicarelli {\it et al.}~\cite{vicarelli_naturemat_2012} have demonstrated room-temperature THz detectors based on antenna-coupled graphene FETs, which exploit the DS mechanism but display also contributions of photo-thermoelectric origin. The plasma waves excited by THz radiation in Ref.~\onlinecite{vicarelli_naturemat_2012} are overdamped and the fabricated detectors, although enabling large area, fast imaging of macroscopic samples, do not yet operate in the aforementioned resonant regime. 

In this Letter we discuss a graphene photodetector in which a dc {\it photocurrent} appears in response to an oscillating field, which is fed between source and drain by the lobes of an antenna.
This coupling geometry between antenna lobes and FETs has already been experimentally implemented~\cite{digaspare_apl_2012} in AlGaAs/InGaAs/AlGaAs heterostructures. Note that in Ref.~\onlinecite{digaspare_apl_2012} the oscillations of the gate-to-channel potential play a crucial role in the rectification of the incoming signal. On the contrary, in our setup, the gate potential is constant (i.e.~it merely fixes the average carrier density) and screens the long-range tail of the carrier-carrier Coulomb interaction. This allows us to use the hydrodynamic theory to describe propagation of plasma waves in the device. Moreover, the setup that we discuss differs from that realized in Ref.~\onlinecite{digaspare_apl_2012} because the rectification which produces the dc photocurrent stems from the intrinsic nonlinear nature of the hydrodynamic theory. The origin of the nonlinear response is thus the same as in the standard DS scheme. Finally, the operating principle of the device proposed in this Letter contrasts the one proposed in Ref.~\onlinecite{ryzhii_wce_2006}, where source and drain are also connected to the lobes of an antenna, but the nonlinear element is a Schottky contact.

The graphene-based device proposed in this work can function as a broadband or resonant detector of THz radiation. In the latter case, the photocurrent as a function of the frequency of the incoming radiation is characterized by sharp peaks at even multiples of the fundamental plasma frequency.
Notably, the photocurrent changes sign at these peaks.

Our analysis relies on the hydrodynamic theory~\cite{bistritzer_prb_2009,rudin_jhses_2011,svintsov_jap_2012,roldan_ssc_2013,tomadin_arxiv_2013}, i.e.~on the combined use of continuity and Euler equations.
The continuity equation reads
\begin{equation}\label{eq:continuity}
\partial_{t}n(x,t) + \partial_{x}\lbrack n(x,t) v(x,t) \rbrack = 0~,
\end{equation}
where $n(x,t)$ is the electron density and $v(x,t)$ is the electron drift velocity.
The $x$ coordinate varies along the transport direction in a field effect transistor geometry in which the source (drain) is placed at $x = 0$ ($x = L$)---see Fig.~\ref{fig:setup}.
We assume that the hydrodynamic variables do not depend on the direction perpendicular to transport.
If the distance $d$ between graphene and the gate is much smaller than the typical wavelength of  plasma oscillations in graphene, the following local relation (known as the ``gradual channel approximation'') between the density and the ``gate-to-channel swing" $U(x,t)$ holds~\cite{tomadin_arxiv_2013}:
\begin{equation}
n(x,t) = \frac{C}{e} U(x,t)~.
\end{equation}
Here, $C = \epsilon_{\rm sub} / (4 \pi d)$ is the geometrical capacitance per unit area, with $\epsilon_{\rm sub}$ the dielectric constant of the insulator separating graphene from the gate, and $e$ is the absolute value of the electron charge.
The gate-to-channel swing can be written in the form $U(x,t) = U_{0} + \delta U(x,t)$, where $U_{0}$ is the gate-to-source potential difference.

We then employ the Euler equation of motion~\cite{tomadin_arxiv_2013}
\begin{equation}\label{eq:euler}
\begin{split}
\partial_{t} v(x,t) & + v(x,t) \partial_{x} v(x,t) = -\frac{e}{m_{\rm c}} \partial_{x} \delta U(x,t) \\
& + \frac{e}{m_{\rm c}} \frac{1}{2 U_{0}} \delta U(x,t) \partial_{x} \delta U(x,t) - \frac{1}{\tau} v(x,t)~.
\end{split}
\end{equation}
Here, $m_{\rm c} = \hbar k_{\rm F} / v_{\rm F}$ is the cyclotron mass, $v_{\rm F} \simeq 1~{\rm nm}/{\rm fs}$ is the Fermi velocity, and $k_{\rm F} = (\pi n_0)^{1/2} = (\pi C U_{0} / e)^{1/2}$ is the Fermi wave number corresponding to the density $n_0 = C U_0/e$ fixed by the gate voltage $U_{0}$.
In writing Eq.~(\ref{eq:euler}) we have neglected contributions~\cite{tomadin_arxiv_2013} due to the pressure and corrections that are important when the Fermi velocity $v_{\rm F}$ is comparable to the plasma wave speed $s = (e U_{0} / m_{\rm c})^{1/2}$.
Note that  Eq.~(\ref{eq:euler}) includes a phenomenological friction term, arising due to scattering of electrons with impurites and phonons, which is proportional to the momentum relaxation rate $\tau^{-1}$.

We consider the setup illustrated in Fig.~\ref{fig:setup}, where the antenna feeds are connected to source and drain.
The incoming radiation, with frequency $\Omega$, generates a potential difference $U_{\rm a}(t) = \epsilon U_{\rm a} \cos{(\Omega t)}$ between the lobes of the antenna. We solve Eqs.~(\ref{eq:continuity}) and~(\ref{eq:euler}) with the following boundary conditions
\begin{equation}\label{eq:boundary}
U(0,t) = U_{0}, \quad
U(L,t) = U_{0} + \epsilon U_{\rm a} \cos{(\Omega t)}~,
\end{equation}
by utilizing a series expansion in the amplitude of the oscillating perturbation. We stress that the dimensionless parameter $\epsilon$ has been introduced to distinguish terms of different order in the series expansion. 
The final result for the dc photocurrent $I$, which is proportional to $U_{\rm a}^2$, is obtained by letting $\epsilon \to 1$. The sign convention for the photocurrent is the following: $I$ is positive when it flows from source to drain.

We start by expanding the hydrodynamic variables in a power series:
\begin{align}
v(x,t) & = \epsilon v_{1}(x,t) + \epsilon^{2} \lbrack \delta v(x) + v_{2}(x,t) \rbrack~, \\
U(x,t) & = U_{0} + \epsilon U_{1}(x,t) + \epsilon^{2} \lbrack \delta U(x) + U_{2}(x,t) \rbrack~.
\end{align}
Here, $U_{n}(x,t)$ and $v_{n}(x,t)$ are periodic functions of time $t$ with frequency $\{n \Omega, n= 0, 1, 2, \dots\}$.
To first order in $\epsilon$, the boundary conditions (\ref{eq:boundary}) become:
\begin{equation}
U_{1}(0,t) = 0,\quad
U_{1}(L,t) = \epsilon U_{\rm a} \cos(\Omega t)~.
\end{equation}
The first-order solutions of Eqs.~(\ref{eq:continuity}) and (\ref{eq:euler}) can be easily determined and read as following:
\begin{equation}\label{eq:solvel1}
v_{1}(x,t) = \frac{U_{\rm a}}{2 U_{0}} \frac{\Omega}{K} \frac{e^{i K x} + e^{-i K x}}{e^{i K L} - e^{-i K L}} e^{-i \Omega t} + \mbox{c.c.}~,
\end{equation}
\begin{equation}\label{eq:solpot1}
U_{1}(x,t) = \frac{U_{\rm a}}{2} \frac{e^{i K x} - e^{-i K x}}{e^{i K L} - e^{-i K L}} e^{-i \Omega t} + \mbox{c.c.}~,
\end{equation}
with $K = (\Omega/s) \sqrt{1 + i/(\Omega \tau)}$.

To second order in $\epsilon$ and after averaging over the period $T = 2\pi / \Omega$ of the incoming radiation, we find:
\begin{equation}\label{eq:continuity2}
\partial_{x} \left [ U_{0} \delta v(x) + \langle U_{1}(x,t) v_{1}(x,t) \rangle_{t} \right ] = 0~,
\end{equation}
\begin{equation}\label{eq:euler2}
\begin{split}
\langle v_{1}(x,t) \partial_{x} v_{1}(x,t) \rangle_{t} & = - \frac{e}{m_{\rm c}} \partial_{x} \delta U(x) - \frac{1}{\tau} \delta v(x) \\
& + \frac{e}{m_{\rm c}} \frac{1}{2 U_{0}} \langle U_{1}(x,t) \partial_{x} U_{1}(x,t) \rangle_{t}~,
\end{split}
\end{equation}
with boundary conditions $\delta U(0) = \delta U(L) = 0$.
In the previous equations $\langle f(t) \rangle_{t} \equiv T^{-1}\int_{0}^{T} dt f(t)$ denotes averaging over time.

\begin{figure}
\includegraphics[width=1.0\linewidth]{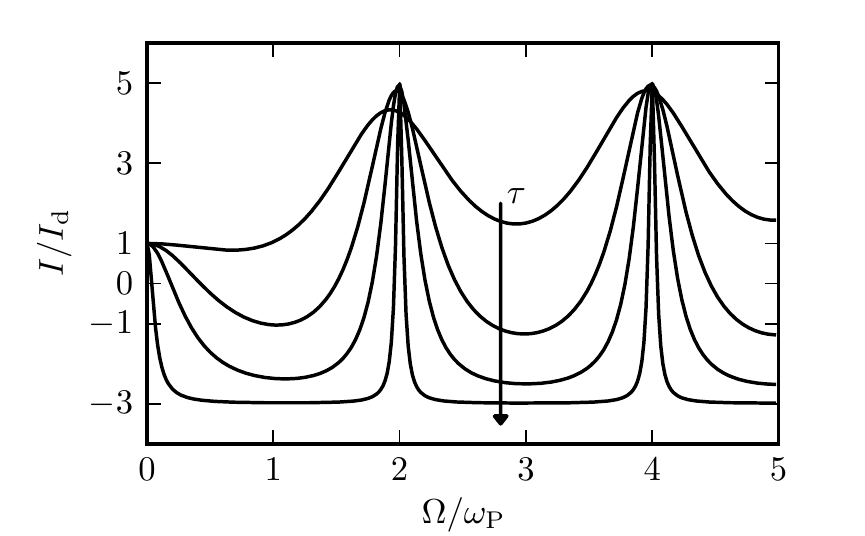}
\caption{\label{fig:current}
The photocurrent $I$ in units of the diffusive current $I_{\rm d}$ [defined in Eq.~(\ref{eq:diffcurr})], which is proportional to $\tau$, is plotted as a function of the ratio between the frequency $\Omega$ of the incoming radiation and the fundamental plasma angular frequency $\omega_{\rm P} = \pi s/(2 L)$.
Different curves correspond to different values of the momentum relaxation time $\tau$.
The values of $\tau$ are $0.5~L / s$, $1.0~L / s$, $2.0~L / s$, and $10.0~L / s$.
The arrow indicates increasing values of $\tau$. }
\end{figure}

We now need to evaluate the current density $J(x,t) = -e n(x,t) v(x,t)$.
The first non-zero contribution to the dc current density is of order $\epsilon^2$ and reads
\begin{equation}\label{eq:curdens2}
J = \langle J(x,t) \rangle_{t} = - \epsilon^{2} C \left [ U_{0} \delta v(x) + \langle U_{1}(x,t) v_{1}(x,t) \rangle_{t} \right ]~.
\end{equation}
Comparing Eq.~(\ref{eq:curdens2}) with Eq.~(\ref{eq:continuity2}), we immediately conclude that $J$ is uniform in space, as expected.
We now solve Eq.~(\ref{eq:curdens2}) for $\delta v(x)$ and substitute the result in Eq.~(\ref{eq:euler2}).
Finally, we integrate the resulting equation in space from $x = 0$ to $x = L$, using that $J$ is uniform.
The final result is:
\begin{equation}\label{eq:quasifinal}
\begin{split}
J & = - \epsilon^{2} \frac{C}{L} \int_{0}^{L} dx \langle U_{1}(x,t) v_{1}(x,t) \rangle_{t} \\
& + \epsilon^{2} \frac{C \tau U_{0}}{L} \int_{0}^{L} dx \langle v_{1}(x,t) \partial_{x} v_{1}(x,t) \rangle_{t} \\
& - \epsilon^{2} \frac{C \tau e}{2 m_{\rm c} L} \int_{0}^{L} dx \langle U_{1}(x,t) \partial_{x} U_{1}(x,t) \rangle_{t}~. 
\end{split}
\end{equation}
The time- and space-integrals in Eq.~(\ref{eq:quasifinal}) can be readily evaluated by employing Eqs.~(\ref{eq:solvel1}) and~(\ref{eq:solpot1}).
The dc photocurrent is given by $I = W_{\rm g} \times \left.J\right|_{\epsilon=1}$, where $W_{\rm g}$ is the width of the device:
\begin{equation}\label{eq:dccurrent}
I = \sigma_{0} \frac{U_{0}}{2} \frac{W_{\rm g}}{L} \left ( \frac{U_{\rm  a}}{2 U_{0}} \right )^{2} \left [ 1 + 2 \beta(\Omega \tau) F(\Omega, \tau) \right ]~,
\end{equation}
where $\beta(x) \equiv 2 x / \sqrt{1 + x^{2}}$ and
\begin{equation}\label{eq:Fmaiuscola}
F(\Omega, \tau) = \frac{\cosh{(2 K_{2} L)} + \cos{(2 K_{1} L)} - 2}{\cosh{(2 K_{2} L)} - \cos{(2 K_{1} L)}}~.
\end{equation}
Here, $K_{1}$ and $K_{2}$ are the real and the imaginary part of the wave number $K$, which depend on $\Omega$ and $\tau$.
In the final expression we have introduced the Drude formula $\sigma_{0} = \sigma_0(U_0) = n_0 e^2 \tau/m_{\rm c}$ for the conductivity of 2D MDFs.
Eq.~(\ref{eq:dccurrent}) is the main result of this Letter.
We note that in the limit $\Omega \ll \tau^{-1}$, which corresponds to the limit of diffusive transport, Eq.~(\ref{eq:dccurrent}) yields $I \to I_{\rm d}$ with
\begin{equation}\label{eq:diffcurr}
I_{\rm d} \equiv \sigma_{0} \frac{U_{0}}{2} \frac{W_{\rm g}}{L} \left ( \frac{U_{\rm a}}{2 U_{0}} \right )^{2}~.
\end{equation}
As expected, the photocurrent $I_{\rm d}$ in the limit of diffusive transport is proportional to the conductivity $\sigma_{0}$, which grows linearly with $\tau$. Note that Eq.~(\ref{eq:diffcurr}) can be written as $I_{\rm d} = W_{\rm g} \sigma_{0} \Delta U_{\rm d} / L$ where $\Delta U_{\rm d} = (U_{\rm a}/2)^2 \sigma^{-1}_0 d\sigma_0/d U_0$ coincides with the well-known 
DS photovoltage that appears when the antenna feeds are connected to gate and source~\cite{vicarelli_naturemat_2012,sakowicz_jap_2011}.
The diffusive result (\ref{eq:diffcurr}) can also be obtained by solving the continuity equation coupled to Ohm's law $J(x,t) = \sigma_0(U(x,t)) \partial_{x} U(x,t)$ where $\sigma_0(U(x,t)) = \left. \sigma_0(U_0)\right |_{U_0 \to U(x,t)}$.

\begin{figure}
\includegraphics[width=1.0\linewidth]{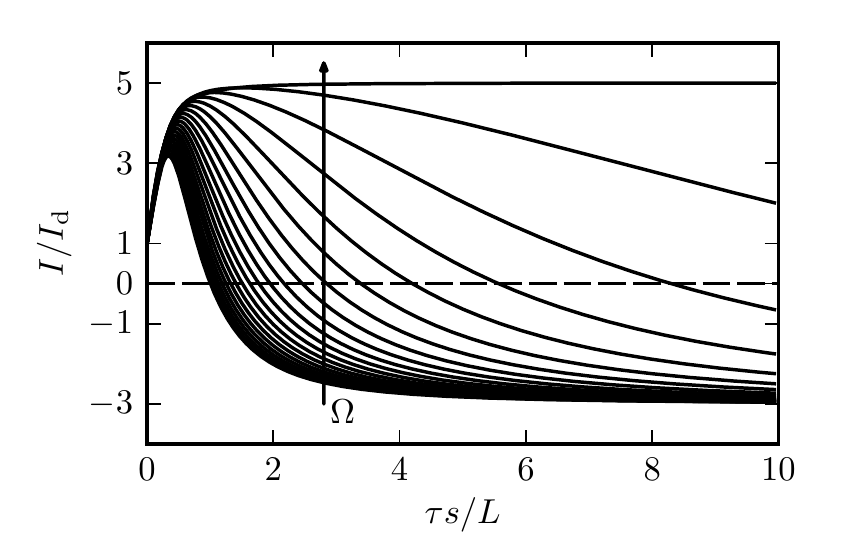}
\caption{\label{fig:steadystate}
The  photocurrent $I$ in units of the diffusive current $I_{\rm d}$ is plotted as a function of the ratio between the momentum relaxation time $\tau$ and $L / s$.
Different curves correspond to different values of the incoming radiation frequency $\Omega$.
The values of $\Omega$ range from $1.5~\omega_{\rm P}$ to $2.0~\omega_{\rm P}$ in steps of $0.025~\omega_{\rm P}$, increasing in the direction indicated by the arrow.}
\end{figure}

Illustrative plots of the photocurrent $I$ (in units of $I_{\rm d} \propto \tau$) are shown in Figs.~\ref{fig:current} and \ref{fig:steadystate}.
In the limit of diffusive transport, the current is positive and frequency independent and its magnitude grows linearly with $\tau$.
In this regime, the device realizes a broadband photodetector.
It is important to notice that a finite dc current is possible because the reflection symmetry $x \to -x$ is explicitly broken by the boundary conditions (\ref{eq:boundary}).
When $\tau$ increases, the current increases (decreases) in the neighborhood of the even (odd) multiples of the fundamental plasma angular frequency $\omega_{\rm P} = \pi s / (2 L)$. For typical device lengths, the plasma-wave frequency $\nu_{\rm P} = \omega_{\rm P} / (2 \pi)$ is in the THz regime~\cite{estimate}. The sign of the current becomes negative in the windows of frequency between even multiples of $\omega_{\rm P}$.
Eventually, for large $\tau$, the current is constant and negative everywhere ($I = - 3 \, I_{\rm d}$) except at even multiples of $\omega_{\rm P}$, where sharp peaks with positive current $I = 5 \, I_{\rm d}$ appear.
In this regime the propagation of plasma waves is ballistic and the device realizes a resonant photodetector.
Tuning of the gate voltage allows to change $\omega_{\rm P}$ and thus to measure $\Omega$ by detecting the sharp switch of the current direction.
From Fig.~\ref{fig:steadystate} we see that the sharp peaks at $\Omega = 2 n \omega_{\rm P}$ are a robust feature, which persists even for $\Omega \tau \gg 1$.

\begin{figure}
\includegraphics[width=1.0\linewidth]{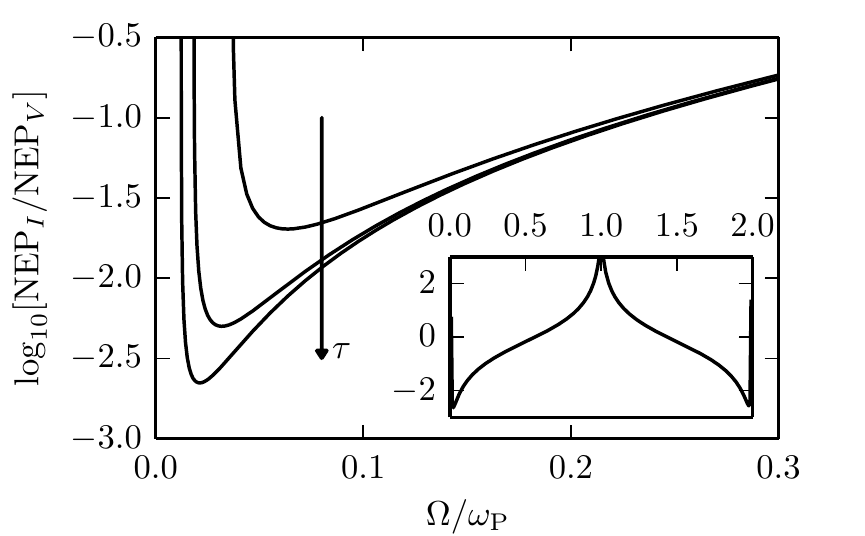}
\caption{\label{fig:nep} The ratio ${\rm NEP}_I/{\rm NEP}_V$ (in logarithmic scale) as from Eq.~(\ref{eq:nep}) is plotted 
as a function of $\Omega$ (in units of $\omega_{\rm P}$). Different curves correspond to different values of the momentum relaxation time $\tau$. 
The values of $\tau$ are $10.0~L / s$, $20.0~L / s$, $30.0~L / s$.
The arrow indicates increasing values of $\tau$. The inset shows the ratio ${\rm NEP}_I/{\rm NEP}_V$ in a larger range of $\Omega$ and for $\tau = 30.0~L/s$. (The axis labels of the inset are not shown since they are the same as in the main panel.) All the results shown in this Figure are restricted to the frequency domain where $I < 0$. }
\end{figure}

We now proceed to compare the performance of our device to that of a standard DS photodetector~\cite{dyakonov_ieee_1996a} in which a dc photovoltage is measured in response to the impinging radiation. Since the current and voltage responsivities of the two setups have different physical dimensions, we compare the noise equivalent power (NEP) per unit bandwidth. The NEP per unit bandwidth has dimensions of ${\rm W}/{\rm Hz}^{1/2}$ in both cases. 

Let us assume that a power $P_{\rm abs}$ is collected by the antenna.
The relation between the potential difference $U_{\rm a}$ at the antenna feeds and $P_{\rm abs}$ can be parameterized by $|U_{\rm a}|^{2} = g P_{\rm abs}$, where $g$ is the so-called coupling efficiency. The current (voltage) responsivity is defined by ${\cal R}_{I} \equiv 
|I| / P_{\rm abs}$ (${\cal R}_{V} \equiv |\Delta U| / P_{\rm abs}$, where $\Delta U$ is the DS photovoltage~\cite{dyakonov_ieee_1996a}). The current noise $I_{\rm n}(\Delta f)$ and the voltage noise $\Delta U_{\rm n}(\Delta f)$ in a bandwidth $\Delta f$ are related by $I_{\rm n}(\Delta f) = \Delta U_{\rm n}(\Delta f) / R$, where $R = L / (W_{\rm g} \sigma_{0})$ is the channel resistance. The NEP in the current (voltage) setup ${\rm NEP}_{I}$ [${\rm NEP}_{V}$] per unit bandwidth is defined by the ratio $I_{\rm n}(\Delta f) / {\cal R}_{I}$ [$V_{\rm n}(\Delta f) / {\cal R}_{V}$].
We therefore find that the NEP per unit bandwidth of our device, measured in units of the NEP per unit bandwidth of a DS photovoltage detector in the usual configuration~\cite{dyakonov_ieee_1996a}, is given by the following ratio:
\begin{equation}\label{eq:nep}
\frac{{\rm NEP}_{I}}{{\rm NEP}_{V}} = \frac{2f(\Omega)}{1 + 2 \beta(\Omega \tau) F(\Omega,\tau)}~,
\end{equation}
where~\cite{dyakonov_ieee_1996a}
\begin{equation}
f(\Omega)  = 1 + \beta(\Omega\tau) - \frac{1+\beta(\Omega\tau)\cos(2K_1 L)}{\sinh^2{(K_2L)}+\cos^2{(K_1 L)}}~,
\end{equation}
and $K_1$, $K_2$, $\beta(\Omega\tau)$, and $F(\Omega,\tau)$ have been defined above. We emphasize that ${\rm NEP}_{I}/{\rm NEP}_{V}$ is independent of the coupling efficiency $g$. 

The ratio in Eq.~(\ref{eq:nep}) is plotted in Fig.~\ref{fig:nep} as a function of $\Omega/\omega_{\rm P}$ and for different values of $\tau$. 
The NEP in our device, i.e. ${\rm NEP}_{I}$, is several orders of magnitude smaller than ${\rm NEP}_{V}$ for frequencies of the incoming radiation $\Omega \ll \omega_{\rm P}$ (main panel in Fig.~\ref{fig:nep}). In this regime, the photocurrent is finite and equal to $-3 I_{\rm d}$ (in the ballistic regime), while the DS photovoltage vanishes in the same limit (see discussion in Sect.~V B of Ref.~\onlinecite{dyakonov_ieee_1996a}). A similar advantage over the DS detection scheme has been noted in Ref.~\onlinecite{lisauskas_ieee_2013}, where the authors propose a photovoltage detection scheme in which the antenna signal is fed to drain and source, as in our setup. From the inset in Fig.~\ref{fig:nep} we see that the ratio of the two NEPs varies over several orders of magnitude and that ${\rm NEP}_{I} \ll {\rm NEP}_{V}$ for $\Omega \sim 2 n \omega_{\rm P}$. The device proposed in this Letter thereby features substantial advantages with respect to the standard DS scheme in terms of signal-to-noise ratio in a wide spectral range. Before concluding, we would like to point out that for typical parameters~\cite{estimate} and sufficiently short devices ($L \lesssim 1~{\rm \mu m}$), the frequency range plotted in the horizontal axis of the main panel of Fig.~\ref{fig:nep} is still in the hundreds of GHz/THz range.

In conclusion, we have shown that a dc photocurrent is generated in a field-effect transistor when the incoming radiation is collimated on the device by connecting the antenna feeds to source and drain.
The generated photocurrent features a peculiar change of sign when the frequency of the radiation matches an even multiple of the fundamental frequency of plasma waves in the channel.
We have carried out a detailed analysis for a graphene field-effect transistor, finding a room-temperature noise equivalent power which, at least in principle, can be comparable to commercially available Terahertz detectors.

\begin{acknowledgments}
We thank Michele Ortolani for useful discussions.
This work was supported by the EC under Graphene Flagship (contract no. CNECT-ICT-604391), 
the Italian Ministry of Education, University, and Research (MIUR) through the program ``FIRB - Futuro in Ricerca 2010'' Grant No.~RBFR10M5BT (``PLASMOGRAPH'') 
and Grant No.~RBFR10LULP (``FRONTERA''),  and the Italian Ministry of Economic Development through the ICE-CRUI project ``TERAGRAPH''.
We have made use of free software (www.gnu.org, www.python.org).
\end{acknowledgments}


\begin{thebibliography}{77}
\bibitem{estimate}
With the choice of parameters $d \sim 100~{\rm nm}$, $k_{\rm F} \sim 0.17~{\rm nm}^{-1}$ (corresponding to a typical doping $n_0 \sim 10^{12}~{\rm cm}^{-2}$) we find $s \sim 8.2~{\rm nm} / {\rm fs}$. In this case, changing $L$ from $300~{\rm nm}$ to $3~\mu{\rm m}$, the plasma-wave frequency in a graphene FET changes from $\nu_{\rm P} \sim 6.8~{\rm THz}$ to $\nu_{\rm P} \sim 680~{\rm GHz}$.
\end{thebibliography}
\end{document}